\documentclass{article}
\usepackage{graphicx}  
\usepackage{amsmath}   
\usepackage[compress]{cite}
\usepackage{amssymb}   
\usepackage{bm} 
\usepackage{dcolumn}
\usepackage{color}
\usepackage{mathrsfs}
\usepackage{amsfonts}
\usepackage{enumerate}
\usepackage{varioref}
\RequirePackage[colorlinks,citecolor=blue,urlcolor=magenta,linkcolor=blue]{hyperref}
\labelformat{section}{Section #1} 
\labelformat{subsection}{Section #1} 
\labelformat{subsubsection}{Section #1}
\labelformat{subsubsubsection}{Section #1}
\labelformat{equation}{Eq.~(#1)} 
\labelformat{figure}{Fig.~#1} 
\labelformat{subfigure}{Fig.~\thefigure#1} 
\labelformat{table}{Tab.~#1} 
\labelformat{appendix}{Appendix #1}

\title{Kinematics of Radion field: A possible source of dark matter}
\author{Sumanta Chakraborty 
\footnote{sumantac.physics@gmail.com;~~sumanta@iucaa.in}\\
{\small {IUCAA, Post Bag 4, Ganeshkhind, Pune University Campus, Pune 411 007, India}}\\
and\\
Soumitra SenGupta
\footnote{tpssg@iacs.res.in}\\
{\small{Department of Theoretical Physics}}\\
{\small {Indian Association for the Cultivation of Science, Kolkata-700032, India}}}
\begin{document}
  
\maketitle
\begin{abstract}
The discrepancy between observed virial and baryonic mass in galaxy clusters have lead to the missing mass problem. To resolve this, a new, non-baryonic matter field, known as dark matter has been invoked. However, till date no possible constituents of the dark matter components are known. This has led to various models, by modifying gravity at large distances to explain the missing mass problem. The modification to gravity appears very naturally when effective field theory on a lower dimensional manifold, embedded in a higher dimensional spacetime is considered. It has been shown that in a scenario with two lower dimensional manifolds separated by a finite distance is capable to address the missing mass problem, which in turn determines the kinematics of the brane separation. Consequences for galactic rotation curves are also described.
\end{abstract}
\section{Introduction}\label{GalBrane_Intro}

Recent astrophysical observations strongly suggest existence of non-baryonic dark matter at the galactic as well as extra-galactic scales (if the dark matter is baryonic in nature, the third peak in the Cosmic Microwave Background power spectrum would have been lower compared to the observed height of the spectrum \cite{Spergel:2006hy}). These observations can be divided into two branches --- (a) behavior of galactic rotation curves and (b) mass discrepancy in clusters of galaxies \cite{Binney:1987}. 

The first one, i.e., rotation curves of spiral galaxies, show clear evidences of problems associated with Newtonian and general relativity prescriptions \cite{Binney:1987,Persic:1995ru,Borriello:2000rv}. In these galaxies neutral hydrogen clouds are observed much beyond the extent of luminous Baryonic matter. In Newtonian description, the equilibrium of these clouds moving in a circular orbit of radius $r$ is obtained through equality of centrifugal and gravitational force. For cloud velocity $v(r)$, the centrifugal force is given by $v^{2}/r$ and the gravitational force by $GM(r)/r^{2}$, where $M(r)$ stands for total gravitational mass within radius $r$. Equating these two will lead to the mass profile of the galaxy as, $M(r)=rv^{2}/G$. This immediately posed serious problem, for at large distances from the center of the galaxy, the velocity remains nearly constant $v\sim 200~\textrm{km/s}$, which suggests that mass inside radius $r$ should increase monotonically with $r$, even though at large distance very 
little luminous matter can be detected \cite{Binney:1987,Persic:1995ru,Borriello:2000rv}. 

The mass discrepancy of galaxy clusters also provides direct hint for existence of dark matter. The mass of galaxy clusters, which are the largest virialized structures in the universe, can be determined in two possible ways --- (i) from the knowledge about motion of the member galaxies one can estimate the virial mass $M_{\rm V}$, secondly, (ii) estimating mass of individual galaxies and then summing over them in order to obtain total baryonic mass $M$. Almost without any exception $M_{\rm V}$ turns out to be much large compared to $M$, typically one has $M_{\rm V}/M \sim 20-30$ \cite{Binney:1987,Persic:1995ru,Borriello:2000rv}. Recently, new methods have been developed to determine the mass of galaxy clusters, these are --- (i) dynamical analysis of hot X-ray emitting gas \cite{Cowie:1986eb} and (ii) gravitational lensing of background galaxies \cite{Grossman:1989} --- these methods also lead to similar results. Thus dynamical mass of galaxy clusters are \emph{always} found to be in excess compared to 
their visible or baryonic mass. This missing mass issue can be explained through postulating that, every galaxy and galaxy cluster is embedded in a halo made up of dark matter. Thus the difference $M_{\rm V}-M$ is originating from the mass of the dark matter halo, the galaxy cluster is embedded in. 

The physical properties and possible candidates for dark matter can be summarized as follows: dark matter is assumed to be non-relativistic (hence cold and pressure-less), interacting only through gravity. Among many others, the most popular choice being weakly interacting massive particles. Among different models, the one with sterile neutrinos (with masses of several keV) has attracted much attention \cite{Albuquerque:2003ei,Viel:2006kd}. Despite of few successes it comes with its own limitations. In the sterile neutrino scenario the X-ray produced from their decay can enhance production of molecular hydrogen and thereby speeding up cooling of gas and early star formation \cite{Biermann:2006bu}. Even after a decade long experimental and observational efforts no non-gravitational signature for the dark matter has ever been found. Thus \emph{a priori} the possibility of breaking down of gravitational theories at galactic scale cannot be excluded \cite{Bekenstein:1984tv,Milgrom:1983ca,
Milgrom:2002tu,Moffat:1995dq,Mannheim:1992vj,Mannheim:1996rv,Milgrom:2003ui,Das:2012ge,Myrzakulov:2015kda}. 

A possible and viable way to modify the behaviour of gravity in our four dimensional spacetime is by introducing extra spatial dimensions. The extra dimensions were first introduced to explain the hierarchy problem (i.e., observed large difference between the weak and Planck energy scales) \cite{ArkaniHamed:1998rs,Antoniadis:1998ig,Rubakov:1983bb}. However the initial works did not incorporate gravity, but used large extra dimensions (and hence large volume factor) to reduce the Planck scale to TeV scale. Introduction of gravity, i.e., warped extra dimensions drastically altered the situation. In \cite{Randall:1999ee} it was first shown that anti-de Sitter solution in higher dimensional spacetime (henceforth referred to as bulk) leads to exponential suppression of the energy scales on the visible four dimensional embedded sub-manifold (called as brane) thereby solving the hierarchy problem. Even though this scenario of warped geometry model solves the hierarchy problem, it also introduces additional 
correction terms to the gravitational field equations, leading to deviations from Einstein's theory at high energy, with interesting cosmological and black hole physics applications \cite{Shiromizu:1999wj,Binetruy:1999ut,Csaki:1999mp,Csaki:1999jh,Chakraborty:2013ipa,Dadhich:2000am,Harko:2004ui,
Chakraborty:2014xla,Chakraborty:2015bja,Chakraborty:2014fva,Chakraborty:2015xja}. This conclusion is not bound to Einstein's gravity alone but holds in higher curvature gravity theories\footnote{In addition to introduction of extra dimensions we could also modify the gravity theory without invoking ghosts, which uniquely fixes the gravitational Lagrangian to be Lanczos-Lovelock Lagrangian. These Lagrangians have special thermodynamic properties and also modifies behaviour of four-dimensional gravity \cite{Chakraborty:2015wma,Chakraborty:2015hna,Padmanabhan:2014jta,Chakraborty:2014rga,Chakraborty:2014joa,Dadhich:2012pd,
Dadhich:2012ma}. However in this work we shall confine ourselves exclusively within the framework of Einstein gravity and shall try to explain the missing mass problem from kinematics of the radion field.} as well \cite{Chakraborty:2014zya,Chakraborty:2014xla,Chakraborty:2015bja}. Since the gravitational field equations get modified due to introduction of extra dimensions it is legitimate to ask, whether it can solve the problem of missing mass in galaxy clusters. Several works in this direction exist and can explain the velocity profile of galaxy clusters. However they emerge through the following setup:
\begin{itemize}

\item obtaining effective gravitational field equations on a lower dimensional hypersurface, starting from the full bulk spacetime, which involves additional contributions from the bulk Weyl tensor. The bulk Weyl tensor in spherically symmetric systems leads to a component behaving as mass and is known as ``dark mass'' (we should emphasize that this notion extends beyond Einstein's gravity and holds for any arbitrary dimensional reduction 
\cite{Chakraborty:2014zya,Chakraborty:2014xla,Chakraborty:2015bja}). It has been shown in \cite{Harko:2007yq} that introduction of the ``dark mass'' term is capable to yield an effect similar to the dark matter. Some related aspects were also explored in \cite{Capozziello:2006uv,Borowiec:2006qr,Pal:2004ii,Boehmer:2007az}, keeping the conclusions unchanged. 

\item In the second approach, the bulk spacetime is always taken to be anti-De Sitter such that bulk Weyl tensor vanishes. Unlike the previous case, which required $S^{1}/Z_{2}$ orbifold symmetry, arbitrary embedding has been considered in \cite{HeydariFard:2008na} following \cite{Maia:2001gq}. This again introduces additional corrections to the gravitational field equations. These additional correction terms in turn lead to the observed virial mass for galaxy clusters. 

\end{itemize}
However all these approaches are valid for a single brane system. In this work we generalize previous results for a two brane system. This approach not only gives a handle on the hierarchy problem at the level of Planck scale but is also capable of explaining the missing mass problem at the scale of galaxy clusters. Moreover, in this setup the additional corrections will depend on the radion field (for a comprehensive discussion see \cite{Chakraborty:2013ipa}), which represents the separation between the two branes. Hence in our setup the missing mass problem for galaxy clusters can also shed some light on the kinematics of the separation between the two branes.

Further the same setup is also shown to explain the observed rotation curves of galaxies as well. Hence both the problems associated with dark matter, namely, the missing mass problem for galaxy clusters and the rotation curves for galaxies can be explained by the two brane system introduced in this work via the kinematics of the radion field.

The paper is organized as follows --- In \ref{Sec_Eff}, after providing a brief review of the setup we have derived effective gravitational field equations on the visible brane which will involve additional correction terms originating from the radion field to modify the gravitational field equations. In \ref{Dark_Virial} we have explored the connection between the radion field, dark matter and the mass profile of galaxy clusters using relativistic Boltzmann equations along with \ref{cluster_App} describing possible applications. Then in \ref{galaxy_dark} we have discussed the effect of our model on the rotation curve of galaxies while \ref{App_Scen} deals with a few applications of our result in various contexts. Finally, we conclude with a discussion on our results. 

Throughout our analysis, we have set the fundamental constant $c$ to unity. All the Greek indices $\mu,\nu,\alpha,\ldots$ run over the brane coordinates. We will also use the standard signature $(-++\ldots)$ for the spacetime metric.
\section{Effective Gravitational Field Equations on the Brane}\label{Sec_Eff}

The most promising candidate for getting effective gravitational field equations on the brane originates from Gauss-Codazzi equation. However these equations are valid on a lower dimensional hypersurface (i.e., on the brane) embedded in a higher dimensional bulk. Hence this works only for a single brane system. But, the brane world model, addressing hierarchy problem requires existence of two branes, where the above method is not applicable. To tackle the problem of two brane system we need to invoke the radion field (i.e., separation between two branes), which has significant role in the effective gravitational field equations. The bulk metric ansatz incorporating the above features takes the following form,
\begin{align}
ds^{2}=e^{2\phi (y,x)}dy^{2}+q_{\mu \nu}(y,x)dx^{\mu}dx^{\nu}
\end{align}
The positive and negative tension branes are located at $y=0$ and $y=y_{0}$ respectively, such that the proper distance between the two branes being given by, $d_{0}(x)=\int _{0}^{y_{0}}dy\exp \phi (x,y)$ and $q_{\mu \nu}$ stands for the induced metric on $y=\textrm{constant}$ hypersurfaces. The effective field equations on the brane depends on the extrinsic curvature, $K_{\mu \nu}=(1/2)\pounds_{n}q_{\mu \nu}$, where the normal to the surface is $n=\exp(-\phi)\partial _{y}$ but also inherits non-local bulk contribution through $E_{\mu \nu}=~^{(5)}C_{\mu \alpha \nu \beta}n^{\alpha}n^{\beta}$, $~^{(5)}C_{\mu \alpha \nu \beta}$ being the bulk Weyl tensor. At first glance it seems that due to non-local bulk effects the effective field equations cannot be solved in closed form, but as we will briefly describe, it can be achieved through radion dynamics and at low energy scales \cite{Shiromizu:2002qr}.

We will now proceed to derive low energy gravitational field equations. As we have already stressed, unless one solves for the non-local effects from the bulk the system of equations would not close. Further it will be assumed that curvature scale on the brane, $L$, is much larger than that of bulk, $\ell$. Then we can expand all the relevant geometrical quantities in terms of the small, dimensionless parameter $\epsilon =(\ell/L)^{2}$. At zeroth order of this expansion, one recovers $^{(0)}q_{\mu \nu}(y,x)=h_{\mu \nu}(x)\exp (-2d(y,x)/\ell)$, while at the first order one has \cite{Shiromizu:2002qr},
\begin{align}
^{(4)}G^{\mu}_{\nu}&=-\frac{2}{\ell}\left(~^{(1)}K^{\mu}_{\nu}-\delta ^{\mu}_{\nu}~^{(1)}K\right)-~^{(1)}E^{\mu}_{\nu}
\label{neweq02}
\\
e^{-\phi}\partial _{y}^{(1)}E_{\mu \nu}&=\frac{2}{\ell} ~^{(1)}E_{\mu \nu}
\label{neweq01}
\\
e^{-\phi}\partial _{y}^{(1)}K^{\mu}_{\nu}&=-\left(D^{\mu}D_{\nu}\phi +D^{\mu}\phi D_{\nu}\phi\right)+\frac{2}{\ell}~^{(1)}K^{\mu}_{\nu}-^{(1)}E^{\mu}_{\nu}
\end{align}
The evolution equations for $^{(1)}E^{\mu}_{\nu}$ and $^{(1)}K^{\mu}_{\nu}$ can be solved as, 
\begin{align}
^{(1)}E^{\mu}_{\nu}&=\exp (4d(y,x)/\ell)\hat{e}^{\mu}_{\nu}(x)
\\
^{(1)}K^{\mu}_{\nu}(y,x)&=\exp(2d(y,x)/\ell)~^{(1)}K^{\mu}_{\nu}(0,x)
-\frac{\ell}{2}\left[1-\exp (-2d(y,x)/\ell) \right] ^{(1)}E^{\mu}_{\nu}(y,x)
\nonumber
\\
&-\left[D^{\mu}D_{\nu}d(y,x)-\frac{1}{\ell}\left(D^{\mu}d D_{\nu}d -\frac{1}{2}\delta ^{\mu}_{\nu}(Dd)^{2}\right)\right]
\end{align}
where, $\hat{e}^{\mu}_{\nu}=h^{\mu \alpha}e_{\alpha \nu}(x)$, with $e_{\alpha \nu}(x)$ being the integration constant of \ref{neweq01} and can be fixed using the junction conditions as \cite{Shiromizu:2002qr},
\begin{align}\label{neweq03}
\frac{\ell}{2}\left[1-\exp(-2d_{0}/\ell)\right]\exp(4d_{0}/\ell)\hat{e}^{\mu}_{\nu}(x)&=-\frac{\kappa ^{2}}{2}\left[\exp(2d_{0}/\ell) T^{({\rm hid})\mu}_{\nu}+T^{({\rm vis})\mu}_{\nu} \right]-\left(D^{\mu}D_{\nu}d_{0}-\delta ^{\mu}_{\nu}D^{2}d_{0}\right)
\nonumber
\\
&+\frac{1}{\ell}\left(D^{\mu}d_{0}D_{\nu}d_{0}+\frac{1}{2}\delta ^{\mu}_{\nu}(Dd_{0})^{2}\right)
\end{align}
where $\kappa ^{2}$ stands for the bulk gravitational constant, $T^{({\rm hid})\mu}_{\nu}$ stands for energy momentum tensor on the hidden (positive tension) brane and $T^{({\rm vis})\mu}_{\nu}$ for the visible (negative tension) brane, respectively. Use of the expressions for $^{(1)}E^{\mu}_{\nu}$ and $^{(1)}K^{\mu}_{\nu}$ in \ref{neweq02} leads to the effective field equations on the visible brane (i.e., the brane on which Planck scale is exponentially suppressed) in this scenario as 
\cite{Shiromizu:2002qr}
\begin{align}\label{Dark_beg}
^{(4)}G^{\mu}_{\nu}&=\frac{\kappa ^{2}}{\ell}\frac{1}{\Phi}T^{(\rm{vis})\mu}_{\nu}+\frac{\kappa ^{2}}{\ell}\frac{\left(1+\Phi \right)^{2}}{\Phi}T^{(\rm{hid})\mu}_{\nu}+\frac{1}{\Phi} \left(D^{\mu}D_{\nu}\Phi -\delta ^{\mu}_{\nu}D^{2}\Phi \right)
\nonumber
\\
&+\frac{\omega (\Phi)}{\Phi ^{2}}\left[D^{\mu}\Phi D_{\nu}\Phi -\frac{1}{2}\delta ^{\mu}_{\nu}\left(D\Phi \right)^{2}\right]
\end{align}
where The scalar field $\Phi (x)$ appearing in the above effective equation is directly connected to the radion field $d_{0}(x)$ (representing proper distance between the branes) such that $\omega (\Phi)$ and $\Phi$ has the following expressions \cite{Shiromizu:2002qr}
\begin{align}\label{Dark_Defn}
\Phi =\exp \left(\frac{2d_{0}}{\ell} \right)-1;\qquad \omega (\Phi)=-\frac{3}{2}\frac{\Phi}{1+\Phi}
\end{align}
We will assume $d_{0}(x)$, the brane separation to be finite and everywhere nonzero. This suggests that $\Phi (x)$ should always be greater than zero and shall never diverge. Finally we also have a differential equation satisfied by $\Phi$ from the trace of \ref{neweq03}, which can be written as \cite{Shiromizu:2002qr}
\begin{align}\label{rad_eq}
D_{\mu}D^{\mu}\Phi =\frac{\kappa ^{2}}{\ell}\frac{1}{2\omega +3}\left(T^{(\rm{vis})}+T^{(\rm{hid})}\right)-\frac{1}{2\omega +3}\dfrac{d\omega}{d\Phi}D_{\mu}\Phi D^{\mu}\Phi
\end{align}
where $\omega (\Phi)$ has been defined in \ref{Dark_Defn} and $T^{\rm vis}$ and $T^{\rm hid}$ stands for the trace of energy momentum tensor on the hidden and visible branes respectively. In the above expressions $D_{\mu}$ stands for the four-dimensional covariant derivative, also $D^{2}\Phi$ stands for $D^{\mu}D_{\mu}\Phi$ and $(D\Phi)^{2}=D_{\mu}\Phi D^{\mu}\Phi$. 

The above effective field equations for gravity have been obtained following \cite{Shiromizu:2002qr}, where no stabilization mechanism for the radion field was proposed. In this work as well we would like to emphasize that, we are working with the radion field in absence of any stabilization mechanism. However as already emphasized in \cite{Shiromizu:2002qr}, in order to provide a possible resolution to the gauge hierarchy problem one requires to stabilize the radion field. Even though we will not explicitly invoke a stabilization mechanism, but will outline how stabilization can be achieved and argue that it will not drastically alter the results.

In such a situation with stabilized radion field, the field $\Phi(x)$ appearing in the above equations can be thought of as fluctuations of the radion field around its stabilized value \cite{Goldberger:1999un}. In particular stabilization of the radion field can be achieved by first introducing a bulk scalar field following \cite{Goldberger:1999uk} and then solving for it. Substitution of the solution in the action and subsequent integration over the extra spatial dimension, leads to a potential for $\Phi$. The same will appear in the above equations through the projection of the bulk energy momentum tensor, which would involve the bulk scalar field and shall lead to an additional potential on the right hand side of the above equations, whose minima would be the stabilized value for $\Phi=\Phi _{c}$. Choosing $\Phi=\Phi _{c}+\Phi(x)$, where $\Phi(x)$ represents small fluctuations around the stabilized value, one ends up with similar equations as above with bulk terms having similar 
contributions as from $T^{\rm (vis)}$ and $T^{\rm (hid)}$ respectively. Thus the final results, to leading order, will remain unaffected by introduction of a stabilization mechanism. Even though the fact that virial mass of galaxy clusters scale with $r$ will hold, the sub-leading correction terms in case of galactic motion will change due to presence of a stabilization mechanism due to appearance of extra bulk inherited terms in the above equations. It would be an interesting exercise to work out the above steps explicitly and obtain the relevant corrections due to stabilization mechanism, which we will pursue elsewhere.

As illustrated above for the two brane system the non-local terms get mapped to the radion field, the separation between the two branes. Hence ultimately one arrives at a system of closed field equations for a two brane system. The field equations as presented in \ref{Dark_beg} are closed since the radion field $\Phi$ satisfies its own field equation \ref{rad_eq}. Hence the problematic non-local terms in a single brane approach gets converted to the radion field in a two brane approach and makes the system of gravitational field equations at low energy closed.

We are mainly interested in spherically symmetric spacetime, in which generically the line element takes the following form:
\begin{align}\label{Dark_Ansatz}
ds^{2}=-e^{\nu}dt^{2}+e^{\lambda}dr^{2}+r^{2}d\Omega ^{2}
\end{align}
This particular form of the metric is used extensively in various physical contexts, for example in obtaining black hole solution, particle orbit, perihelion precession of planetary orbits, bending of light and in various other astrophysical phenomenon \cite{Chakraborty:2012sd,Chakraborty:2011uj,Chakraborty:2012mz}. Given this metric ansatz we can compute all the derivatives of the scalar field and being a static situation, the brane separation is assumed to depend on radial coordinate only. Thus we will only have terms involving derivative with respect to $r$ (these will be denoted by prime). First we can rewrite the scalar field equation, which will be a differential equation for $\Phi$. We will also assume that there is no matter on the hidden brane, but only on the visible brane, which is assumed to be perfect fluid. Thus on the visible brane we have energy momentum tensor to be, $T^{\nu({\rm vis})}_{\mu}=\textrm{diag} (-\rho,p,p_{\perp},p_{\perp})$, with the trace being given by, $T=-\rho +p+2p_{\perp}$.
 From now on we will remove the label `vis' from the energy momentum tensor, since only on the visible brane energy momentum tensor is non-zero. With these inputs and the above spherically symmetric metric ansatz we obtain the scalar field equation as,
\begin{align}
\partial _{r}^{2}\Phi +\frac{2}{r}\partial _{r}\Phi +\left(\frac{\nu '-\lambda '}{2}\right)\partial _{r}\Phi=\frac{\kappa ^{2}}{\ell}\frac{1+\Phi}{3}T^{(\rm{vis})}e^{\lambda}+\frac{1}{2(1+\Phi)}\left(\partial _{r}\Phi \right)^{2}
\end{align}
Having derived the scalar field equation, next we need to obtain the field equations for gravity with the metric ansatz given by \ref{Dark_Ansatz}. These will be differential equations for $\nu (r)$ and $\mu (r)$ respectively. We can separate out the time-time component, radial component and transverse components leading to
\begin{align}
-e^{-\lambda}\left(\frac{1}{r^{2}}-\frac{\lambda '}{r}\right)+\frac{1}{r^{2}}&=\frac{\kappa ^{2}}{\ell}\frac{\rho +\rho _{0}}{\Phi}+e^{-\lambda}\frac{\nu '}{2}\frac{\Phi '}{\Phi}+\frac{\kappa ^{2}}{\ell}\frac{1+\Phi}{3\Phi}T-\frac{e^{-\lambda}\Phi '^{2}}{4\Phi(1+\Phi)}
\label{GB_Eq01a}
\\
e^{-\lambda}\left(\frac{\nu '}{r}+\frac{1}{r^{2}}\right)-\frac{1}{r^{2}}&=\frac{\kappa ^{2}}{\ell}\frac{p-\rho _{0}}{\Phi}-\frac{2}{r}e^{-\lambda}\frac{\Phi '}{\Phi}-e^{-\lambda} \frac{\nu '}{2}\frac{\Phi '}{\Phi}-\frac{3}{4}\frac{\Phi '^{2}}{\Phi (1+\Phi)}e^{-\lambda}
\label{GB_Eq01b}
\\
e^{-\lambda}\left(\nu ''+\frac{\nu '^{2}}{2}+\frac{\nu '-\lambda '}{r}-\frac{\nu '\lambda '}{2}\right)&=\frac{\kappa ^{2}}{\ell \Phi}2\left(p_{\perp}-\rho _{0} \right)+\frac{2}{r}e^{-\lambda}\frac{\Phi '}{\Phi}-\frac{2\kappa ^{2}}{3\ell}\frac{1+\Phi}{\Phi}T+\frac{1}{2}\frac{e^{-\lambda}\Phi '^{2}}{\Phi (1+\Phi)} 
\label{GB_Eq01c}
\end{align}
where primes denote derivatives with respect to radial coordinate. In the above field equations along with the perfect fluid, we have contributions from the brane cosmological constant. Here we have inserted a brane energy density $\rho _{0}$, where $\rho _{0}$ and brane cosmological constant is related via $\rho _{0}=\Lambda /8\pi G$. Here $G$ is the four dimensional gravitational constant. Finally, we have contribution from the radion field itself, since it appears on the right hand side of gravitational field equations. Having derived the field equations we will now proceed to determine the effect of the radion field on the kinematics of galaxy clusters and hence its implications for the missing mass problem.  
\section{Virial Theorem in Galaxy Clusters, Kinematics of The Radion Field and Dark Matter}\label{Dark_Virial}

It is well known that the galaxy clusters are the largest virialized systems in the universe \cite{Binney:1987}. We will further assume them to be isolated, spherically symmetric systems such that the spacetime metric near them can be presented by the ansatz in \ref{Dark_Ansatz}. Galaxies within the galaxy cluster are treated as identical, point particles satisfying general relativistic collision-less Boltzmann equation. 

The Boltzmann equation requires setting up appropriate phase space for a multi-particle system along with the corresponding distribution function $f(x,p)$, where $x$ is the position of the particles in the spacetime manifold with its four-momentum $p\in T_{x}$, where $T_{x}$ is the tangent space at $x$. Further the distribution function is assumed to be continuous, non-negative and describing a state of the system. The distribution function is defined on the phase space, yielding the number $dN$ of the particles of the system, within a volume $dV$ located at $x$ and have four-momentum $p$ within a three surface element $d\overrightarrow{p}$ in momentum space. All the observables can be constructed out of various moments of the distribution function. Further details can be found in \cite{Harko:2007yq}.

For the static and spherically symmetric line element as in \ref{Dark_Ansatz} the distribution function can depend on the radial coordinate only and hence the relativistic Boltzmann equation reduces to the following form \cite{Harko:2007yq},
\begin{align}
u_{r}\frac{\partial f}{\partial r}-\left(\frac{1}{2}u_{t}^{2}\frac{\partial \nu}{\partial r}-\frac{u_{\theta}^{2}+u_{\phi}^{2}}{r}\right)\frac{\partial f}{\partial u_{r}}&-\frac{1}{r}u_{r}\left(u_{\theta}\frac{\partial f}{\partial u_{\theta}}+u_{\phi}\frac{\partial f}{\partial u_{\phi}}\right)
\nonumber
\\
&-\frac{1}{r}e^{\lambda /2}u_{\phi}\cot \theta \left(u_{\theta}\frac{\partial f}{\partial u_{\phi}}-u_{\phi}\frac{\partial f}{\partial u_{\theta}}\right)=0
\end{align}
The spherical symmetry of the problem requires the coefficient of $\cot \theta$ to identically vanish. Hence the distribution function can be a function of $r$, $u_{r}$ and $u_{\theta}^{2}+u_{\phi}^{2}$ only. Multiplying the above equation by $mu_{r}du$, where $m$ stands for galaxy mass and $du$ is the velocity space element we find after integrating over the cluster \cite{Harko:2007yq},
\begin{align}\label{GB_Eq03}
-\int _{0}^{R}4\pi \rho \left[\langle u_{r}^{2}\rangle +\langle u_{\theta}^{2}\rangle +\langle u_{\phi}^{2}\rangle \right]r^{2}dr
+\frac{1}{2}\int _{0}^{R} 4\pi r^{3}\rho \left[\langle u_{t}^{2}\rangle +\langle u_{r}^{2}\rangle \right] \frac{\partial \nu}{\partial r}dr =0
\end{align}
where $R$ stands for the radius of the galaxy cluster. Using the distribution function, the energy momentum tensor of the matter becomes,
\begin{align}
T_{ab}=\int fmu_{a}u_{b}du
\end{align}
which leads to the following expressions for energy density and pressure as,
\begin{align}
\rho _{\rm eff}=\rho \langle u_{t}^{2}\rangle;\qquad p_{\rm eff}^{(r)}=\rho \langle u_{r}^{2}\rangle;\qquad p_{\rm eff}^{(\perp)}=\rho \langle u_{\theta}^{2}\rangle =\rho \langle u_{\phi}^{2}\rangle 
\end{align}
Using these expressions for energy density and pressure in the gravitational field equations presented in \ref{GB_Eq01a}, \ref{GB_Eq01b} and \ref{GB_Eq01c} and finally adding all of them together we arrived at
\begin{align}\label{GB_Eq02}
e^{-\lambda}\left(\nu ''+2\frac{\nu '}{r}+\frac{\nu '^{2}}{2}-\frac{\nu '\lambda '}{2}\right)&=\frac{\kappa ^{2}}{\ell \Phi}\left(\rho _{\rm eff} +p_{\rm eff}^{(r)}+2p^{(\perp)}_{\rm eff}\right)-\frac{1}{2}\frac{e^{-\lambda}\Phi '^{2}}{\Phi (1+\Phi)}
\nonumber
\\
&-\frac{\kappa ^{2}}{3\ell}\frac{1+\Phi}{\Phi}\left(-\rho _{\rm eff}+p^{(r)}_{\rm eff}+p^{(\perp)}_{\rm eff}-4\rho _{0}\right)-\frac{\kappa ^{2}}{\ell \Phi}\rho _{0}
\end{align}
To obtain \ref{GB_Eq02}, we have used the expression for trace of the energy momentum tensor. We also remember that $\rho _{0}$ stands for vacuum energy density. At this stage it is useful to introduce certain assumptions, since  actually we are interested in a post-Newtonian formulation of the effective gravitational field equations. The two assumptions are --- (a) $\nu$ and $\lambda$ are small so that any quadratic expressions constructed out of them can be neglected in comparison to the linear one. Secondly, (b) the velocity of the galaxies are assumed to be much smaller compared to the velocity of light, which suggests, $\langle u_{r}^{2}\rangle$, $\langle u_{\theta}^{2}\rangle$, $\langle u_{\phi}^{2}\rangle$ $\ll$ $\langle u_{t}^{2}\rangle$. This in turn implies $\rho _{\rm eff} \gg p_{\rm eff}^{(r)},p^{(\perp)}_{\rm eff}$ such that all the pressure terms can be neglected in comparison to the energy density. Applying all these approximation schemes \ref{GB_Eq02} can be rewritten as,
\begin{align}\label{GB_Eq04}
\frac{1}{2r^{2}}\dfrac{\partial}{\partial r}\left(r^{2}\nu '\right)=\frac{\kappa ^{2}}{6\ell}\rho +\frac{2\kappa ^{2}}{3\ell}\rho _{0}-\frac{1}{4}\frac{\Phi '^{2}}{\Phi (1+\Phi)}+\frac{2\kappa ^{2}\rho}{3\Phi \ell}-\frac{\kappa ^{2}\rho _{0}}{3\ell \Phi}
\end{align}
We can also perform the same schemes of approximation to \ref{GB_Eq03}, which leads to, 
\begin{align}
-2K+\frac{1}{2}\int _{0}^{R}4\pi r^{3}\rho \nu ' dr=0
\end{align}
where $K$ stands for the total kinetic energy of the galaxies within the galaxy cluster and has the following expression,
\begin{align}
K=\int _{0}^{R}dr~4\pi r^{2}\rho \left[\frac{1}{2}\left\lbrace\langle u_{r}^{2}\rangle +\langle u_{\theta}^{2}\rangle +\langle u_{\phi}^{2}\rangle \right\rbrace \right]
\end{align}
The mass within a small volume of radial extent $dr$ has the expression $dM(r)=4\pi r^{2}\rho dr$, where in this and subsequent expressions $\rho$ will indicate $\rho (r)$. Thus total mass of the system can be given by integral of $dM(r)$ over the full size of the galaxy. The main contribution comes from mass of intra-cluster gas and stars along with other particles, e.g., massive neutrinos. We can also define the gravitational potential energy $\Omega$ of the cluster as,
\begin{align}
\Omega =-\int _{0}^{R}\frac{GM(r)}{r}dM(r)
\end{align}
Finally multiplying \ref{GB_Eq04} by $r^{2}$ and integrating from $0$ to $r$, we arrive at,
\begin{align}\label{GB_Eq05}
\frac{1}{2}r^{2}\dfrac{\partial \nu}{\partial r}&=\frac{\kappa ^{2}}{6\ell}\int ^{r}_{0}r^{2}\rho (r)dr +\frac{2\kappa ^{2}\rho _{0}}{3\ell}\int _{0}^{r}r^{2}dr+\frac{\kappa ^{2}}{4\pi \ell}M_{\Phi}(r)
\end{align}
where we have defined:
\begin{align}\label{radion_mass}
M_{\Phi}(r)=\int ^{r}_{0}dr 4\pi r^{2}\left(-\frac{\ell}{4\kappa ^{2}}\frac{\Phi '^{2}}{\Phi (1+\Phi)}+\frac{2\rho}{3\Phi}-\frac{\rho _{0}}{3\Phi}\right)
\end{align}
This object captures all the effect of the radion field on the gravitational mass distribution of galaxy clusters and thus may be called as the ``radion mass''. Note that the ``radion mass'' defined in this work, is a completely different construct compared to the ``dark mass'' used in the literature. The dark mass appears from non-local effects of the bulk, specifically through the bulk Weyl tensor in the effective field equation formalism. However, in this work, we have used effective equation formalism for a two brane system as developed in \cite{Shiromizu:2002qr}, where the correction to gravitational field equations originate from radion dynamics. Pursuing this effective equations further, through virial theorem we have shown that the effect of radion dynamics can be summarized by introducing a radion mass as in \ref{radion_mass}. Hence conceptually and structurally the dark mass of \cite{Harko:2007yq} is completely different from our ``radion mass''.

Further, the total baryonic mass of the galaxy cluster within a radius $r$ can be obtained by integrating the energy density over the size of the galaxy cluster, which leads to, $M(r)=4\pi \int ^{r}_{0}r^{2}\rho (r)dr$, using which we finally arrive at the following form for \ref{GB_Eq05}:
\begin{align}
\frac{1}{2}r^{2}\dfrac{\partial \nu}{\partial r}&=\frac{\kappa ^{2}}{6\ell}\frac{M(r)}{4\pi}+\frac{2\kappa ^{2}\Lambda}{3\ell}\frac{r^{3}}{3}+\frac{\kappa ^{2}}{4\pi \ell}M_{\Phi}(r)
\end{align}
Earlier we have defined the gravitational potential associated with $M$, the baryonic mass. We can define an identical object using the radion mass as well, leading to an potential term $\Omega _{\Phi}$. Given the potentials we can introduce three radius --- (a) $R_{V}$, the virial radius, obtained using total baryonic potential and baryonic mass, (b) $R_{I}$, the inertia radius, obtained from moment of inertia of the galaxy cluster and finally (c) $R_{\Phi}$, the radion radius obtained from the radion mass (for detailed expressions see \ref{APP_01}). Using these expressions and the definition for virial mass, $M_{V}=\sqrt{2KR_{V}/G}$, yields the following expression,
\begin{align}\label{GB_Final}
\frac{M_{V}}{M}=\sqrt{\frac{\kappa ^{2}}{24\pi G\ell}+\frac{2\kappa ^{2}\rho _{0}}{9\ell G}\frac{R_{V}R_{I}^{2}}{M}+\frac{\kappa ^{2}}{4\pi G\ell}\frac{R_{V}}{R_{\Phi}}\frac{M_{\Phi}^{2}}{M^{2}}}
\end{align}
For most of the clusters, the virial mass $M_{V}$ is three times compared to the baryonic mass $M$ and thus for all practical purposes the first term inside the square root, which is of order unity can be neglected with respect to the other two. The second term yields the contribution from the brane cosmological constant, which is several orders of magnitude smaller compared to the observed mass and thus can also be neglected. Finally, the virial mass turns out to be,
\begin{align}
\frac{M_{V}}{M}\approx \frac{M_{\Phi}}{M}\sqrt{\frac{\kappa ^{2}}{4\pi G\ell}\frac{R_{V}}{R_{\Phi}}}
\end{align}
Among the various terms in the above expression, virial mass $M_{V}$ is determined from the study of velocity dispersion of galaxies within the cluster and is much large than the visible mass. The above expression shows that if the radion field kinematics is such that $M_{\rm tot}$ is equal to $M_{\Phi}$. Then that in turn will lead to the correct virial mass of the galaxy clusters. The effect of radion field and hence of extra dimension can also be probed through gravitational lensing. 

To see that, let us explore the differential equation for $\Phi$, which has not yet been considered. Solving that will lead to some leading order behaviour of the radion field $\Phi$, which in turn would affect $M_{\Phi}$. Thus crucial thing is whether $M_{\Phi}$ behaves as $r$ at large distance from the core of the cluster. In which case from the above equation, we readily observe that the galaxy virial mass would also scale as $M_{V}\sim r$ explaining the issue of dark matter and galaxy rotation curve. To answer all these let us start by using the differential equation for $\Phi$. There we will work under same approximation schemes, i.e., will be neglecting all the quadratic terms, e.g., $\nu '\Phi '$, $\Phi '^{2}$, will set $e^{\lambda}\sim 1$ and shall neglect vacuum energy contribution $\rho _{0}$ to obtain (for general expression see \ref{APP_01}), 
\begin{align}
\Phi ''+\frac{2}{r}\Phi '=-\frac{\kappa ^{2}}{3\ell}\left(1+\Phi \right)\rho 
\end{align}
Multiplying both sides by $r^{2}$ and integrating twice we obtain (noting that $\Phi '^{2}$ should not contribute)
\begin{align}\label{Rad_Final}
\Phi =-\frac{\kappa ^{2}}{12\pi \ell}\int dr~\frac{M(r)}{r^{2}}
\end{align}
Here $M(r)$ stands for the mass of the baryonic matter within radius $r$ and we know from observations that the density of the baryonic matter falls as $\rho _{c}(r_{c}/r)^{\beta}$, where $3>\beta >2$ and $r_{c}$ stands for the core radius of the cluster. Thus it is straightforward to compute the mass profile, which goes as $\sim r^{3-\beta}$, except for some constant contribution. Hence finally after integration we obtain the radion field to vary with the radial distance as $r^{2-\beta}$. However note that the mass of the radion field, i.e., $M_{\Phi}$ under these approximations (matter is non-relativistic and field is weak) can be obtained as,
\begin{align}\label{rad_mass}
M_{\Phi}(r)=\int ^{r}_{0}dr 4\pi r^{2}\frac{2\rho}{3\Phi}=8\pi(\beta -2)(3-\beta)\frac{\ell}{\kappa ^{2}}r
\end{align}
Thus the radion mass indeed scales linearly with radial distance which would correctly reproduce the observed virial mass of the galaxy cluster. Due to the linear nature of the virial mass, the velocity profile does not die down at large $r$ as expected. Hence the radion field kinematics can explain the kinematics of the galaxy cluster very well and thus the missing mass problem can be described without invoking any additional matter component. 

Before concluding the section, let us briefly mention about the connection of the above formalism with the gauge hierarchy problem. The separation between the two branes is denoted by $d$, which varies with the radion field $\Phi$, logarithmically (see \ref{Dark_Defn}). The radion field except for a constant contribution varies weakly with radial distance and hence leads to very small corrections to the distance $d$ between the branes. Thus the graviton mass scale for the visible brane will be suppressed by a similar exponential factor as in the original scenario of Randall and Sundrum \cite{Randall:1999ee,Lykken:1999nb}, leading to a possible resolution of the gauge hierarchy problem. Thus as advertised earlier, existence of an extra spatial dimension leads to a radion field, producing a possible explanation for the dark matter in galaxy clusters along with solving the gauge hierarchy problem.
\section{Application: Cluster mass profiles}\label{cluster_App}

In the previous section we have discussed galaxy clusters by assuming them to be bound gravitational systems, with approximate spherical symmetry and virialized, i.e., in hydrostatic equilibrium. With these reasonable set of assumptions we have shown that, the mass of clusters receive additional contribution from the kinematics of the radion field and provides an alternative to the missing mass problem. In this section we will discuss one application of the above formalism, namely the mass profile of galaxy clusters and possible experimental consequences. We again start from collisionless Boltzmann equation in spherical symmetry and in hydrostatic equilibrium to read,
\begin{align}\label{neweq05}
\frac{d}{dr}\left[\rho _{\rm gas}(r)\sigma _{r}^{2}\right]+\frac{2\rho _{\rm gas}(r)}{r}\left(\sigma _{r}^{2}-\sigma _{\theta ,\phi}^{2}\right)=-\rho _{\rm gas}(r)\frac{dV(r)}{dr}
\end{align}
Here $V(r)$ stands for the gravitational potential of the cluster, $\sigma _{r}$ and $\sigma _{\theta ,\phi}$ are the mass weighted velocity dispersions in the radial and tangential directions, respectively, with $\rho _{\rm gas}$ being the gas density. For spherically symmetric systems, $\sigma _{r}=\sigma _{\theta,\phi}$ and the pressure profile becomes, $P(r)=\sigma _{r}^{2}\rho _{\rm gas}(r)$. Further if the velocity dispersion is assumed to have originated from thermal fluctuations, for a gas sphere with temperature profile $T(r)$, the velocity dispersion becomes, $\sigma _{r}^{2}=k_{B}T(r)/\mu m_{p}$, where $k_{B}$ is the Boltzmann constant, $\mu \simeq 0.609$ is the mean mass and $m_{p}$ being the proton mass. Thus \ref{neweq05} can be rewritten as,
\begin{align}
\frac{d}{dr}\left[\frac{k_{B}T(r)}{\mu m_{p}}\rho _{\rm gas}(r)\right]=-\rho _{\rm gas}(r)\frac{dV(r)}{dr}
\end{align}
The potential can be divided into two parts, the Newtonian potential and the potential due to radion field. As multiplied by $(4/3)r^{2}/G$, the Newtonian potential leads to the Newtonian mass $M_{N}$, which includes mass of gas, galaxies and in particular of the CD galaxies. Thus finally we obtain the mass profile of a virialized galaxy cluster to be,
\begin{align}
M_{\rm N}(r)+\frac{4}{3G}r^{2}\frac{dV_{\Phi}}{dr}=-\frac{4}{3}\left(\frac{k_{B}T(r)}{\mu m_{p}G}\right)r \left(\frac{d \ln \rho _{\rm gas}(r)}{d\ln r}+\frac{d \ln T(r)}{d \ln r}\right)
\end{align}
Thus one needs two experimental input, the observed gas density profile, $\rho _{\rm gas}$ and the observed temperature profile $T(r)$. Gas density can be obtained from the characteristic properties of observed X-ray surface brightness profiles, similarly from X-ray spectral analysis one obtains the radial profile of temperature. Thus from X-ray analysis one can model the galaxy distribution and obtain the baryonic contribution to the mass of the galaxy cluster. From the difference between virial mass and the above estimate one can obtain the contribution due to radion field. At the leading order the radion mass scales linearly with the radial distance with its coefficients being $\mathcal{O}(\ell/\kappa ^{2})$. Thus an estimate of the radion mass will lead to a possible value for $\ell/\kappa ^{2}$. Assuming the bulk gravitational constant to be at the Planck scale one can possibly constrain the bulk curvature scale. 
\section{Effect on galaxy rotation curves}\label{galaxy_dark}

Having described a possible resolution of the missing mass problem in connection with galaxy clusters, let us now concentrate on the rotation curves of galaxies. To perform the same we would invoke some general Lie groups of transformation on a vacuum brane spacetime. In particular we will assume the metric to be static and spherically symmetric (i.e., expressed as in \ref{Dark_Ansatz}), such that, $\pounds _{\xi}g_{\mu \nu}=\psi (r)g_{\mu \nu}$, where the vector field $\xi ^{\mu}$ can be time dependent. These are known as conformally symmetric vacuum brane model and we consider angular velocity of a test particle in visible (i.e., negative tension) brane, which can be determined in terms of the conformal factor $\psi (r)$. The above essentially amounts to the assumption that each brane is conformally mapped onto itself along the vector field $\xi ^{\mu}$ \cite{Maartens:1989ay,Mak:2003kw,Mak:2004hv}. It turns out that both the metric and the vector field $\xi ^{\mu}$ has the following 
expressions upon solving the relation $\pounds _{\xi}g_{\mu \nu}=\psi (r)g_{\mu \nu}$ as,
\begin{align}
\xi ^{\mu}&=\left(\frac{1}{2}\frac{k}{B}t,\frac{r\psi (r)}{2},0,0\right)
\\
e^{-\lambda}&=\psi ^{2}/B^{2};\qquad e^{\nu}=C^{2}r^{2}\exp \left(-2\frac{k}{B}\int \frac{dr}{r\psi}\right)
\label{newadd01}
\end{align}
where $k$ is a separation constant and $B$ and $C$ are integration constants. Substitution of these metric functions in the gravitational field equations presented in \ref{Dark_beg} leads to,
\begin{align}
-\frac{\psi ^{2}}{B^{2}}\left(\frac{1}{r^{2}}+\frac{2}{r}\frac{\psi '}{\psi}\right)+\frac{1}{r^{2}}&=-\frac{e^{-\lambda}\Phi '^{2}}{4\Phi(1+\Phi)}
\label{neweq04a}
\\
\frac{\psi ^{2}}{B^{2}}\left(\frac{3}{r^{2}}-2\frac{k}{B}\frac{1}{r^{2}\psi} \right)-\frac{1}{r^{2}}&=-\frac{3}{4}\frac{\Phi '^{2}}{\Phi (1+\Phi)}e^{-\lambda}
\label{neweq04b}
\\
\frac{\psi ^{2}}{B^{2}}\left(2\frac{\psi '}{r\psi}-2\frac{k}{B}\frac{1}{r^{2}\psi}+\frac{k^{2}}{B^{2}}\frac{1}{r^{2}\psi ^{2}}+\frac{1}{r^{2}}\right)&=\frac{1}{4}\frac{e^{-\lambda}\Phi '^{2}}{\Phi (1+\Phi)} 
\label{neweq04c}
\end{align}
where `prime' denotes differentiation by radial coordinate $r$. Multiplying \ref{neweq04c} by $2$ and adding it to \ref{neweq04b} one can readily equate it to \ref{neweq04a} resulting into the following differential equation satisfied by $\psi (r)$,
\begin{align}
3r\psi \psi '+3\psi ^{2}-3\frac{k}{B}\psi +\frac{k^{2}}{B^{2}}-B^{2}=0
\label{newadd03}
\end{align}
The above differential equation can be readily solved, yielding $r=r(\psi)$ \cite{Maartens:1989ay,Mak:2003kw,Mak:2004hv}. However the solution depends on the mutual dependence of $k$ on $B$. We will use galaxy rotation curves as the benchmark to determine the region of interest in the $(k,B)$ plane. In connection to rotation curves, the motion of a particle on a circular orbit and its tangential velocity is of importance. For the static and spherically symmetric spacetime the tangential velocity of a particle in circular orbit corresponds to,
\begin{align}
v_{\rm tg}^{2}=\frac{r\nu '}{2}=1-\frac{k}{B}\frac{1}{\psi}
\label{newadd02}
\end{align}
where the last equality follows from \ref{newadd01}. The above relation further shows the fact that rotational velocity is determined by $g_{rr}$ component alone. Since $v_{\rm tg}$ is determined by $\psi$, it is possible to write all the expressions derived earlier in terms of the tangential velocity, e.g., $\exp(\lambda)=(B^{4}/k^{2})(1-v_{\rm tg}^{2})^{2}$. From \ref{newadd02} it is clear that asymptotic limits exist only if $k\in (-2B^{2},2B^{2})$ \cite{Mak:2004hv}. In this case the solution to \ref{newadd03} corresponds to,
\begin{align}
r^{2}=R_{0}^{2}\frac{\left(\frac{|\psi -\psi _{2}|}{|\psi -\psi _{1}|}\right)^{m}}{|3\psi ^{2}-3\frac{k}{B}\psi +\frac{k^{2}}{B^{2}}-B^{2}|};\qquad
\psi _{1,2}=\frac{3\frac{k}{B}\pm \sqrt{12B^{2}-3\frac{k^{2}}{B^{2}}}}{6};\qquad
m=\frac{3k}{B\sqrt{12B^{2}-3\frac{k^{2}}{B^{2}}}}
\end{align}
Use of this solution leads to the following asymptotic expression for the tangential velocity,
\begin{align}
v_{{\rm tg},\infty}=\sqrt{1-\frac{6k}{3k+\sqrt{12B^{4}-3k^{2}}}}
\end{align}
Note that for the following choices, $B=1.00000034$ and $k=0.9$ the limiting tangential velocity is given by $v_{{\rm tg},\infty} \sim 216.3 \textrm{km/s}$, which is of the same order as the observed galactic rotational velocities. Thus behavior of all the metric coefficients in the solutions depend on two arbitrary constants of integration, namely, $k$ and $B$. In order to obtain numerical estimate for these parameters we assume that there exist some radius $r_{0}$ beyond which baryonic matter density $\rho _{B}$ is negligible. Requiring $\exp(\lambda)=1-(2GM_{B}/r_{0})$, with $M_{B}=4\pi \int _{0}^{r_{0}}dr r^{2}\rho _{B}$, we readily obtain,
\begin{align}
\frac{k^{2}}{B^{4}}=\left(1-\frac{2GM_{B}}{r_{0}}\right)\left[1-v_{\rm tg}^{2}(r_{0})\right]^{2}
\end{align}
Hence the ratio $k^{2}/B^{4}$ can be determined observationally through the tangential velocity. It follows that around and outside $r_{0}$ the radion field will dominate and hence one can introduce a ``radion mass'' in an identical manner. This on use of conformal symmetry and the ratio $k^{2}/B^{4}$ from the above equation immediately reads,
\begin{align}
M_{\Phi}(r)&=\int dr 4\pi r^{2}\left(\frac{\ell}{\kappa ^{2}}\right)\left[\frac{1}{r^{2}}-e^{-\lambda}\left(\frac{1}{r^{2}}-\frac{\lambda '}{r}\right)\right]
\nonumber
\\
&=4\pi \frac{\ell}{\kappa ^{2}}\left[r-\int dr r^{2}\frac{\psi ^{2}(r)}{B^{2}}\left(\frac{1}{r^{2}}-\frac{\lambda'}{r} \right)\right]
\nonumber
\\
&=\frac{4\pi \ell r}{\kappa ^{2}}\left[1-\left(1-\frac{2GM_{B}}{r_{0}}\right)\frac{1-v_{\rm tg}^{2}(r_{0})}{1-v_{\rm tg}^{2}(r)} \right]
\end{align}
However the tangential velocity $v_{\rm tg}$ is non-relativistic, i.e., much smaller than unity (in $c=1$ units) and hence the radion mass turns out have a scaling relation, 
\begin{align}
M_{\Phi}(r)=\frac{8\pi G\ell}{\kappa ^{2}}M_{B}\frac{r}{r_{0}}
\end{align}
The above result explicitly shows that, the ``radion mass'', will scale linearly with the radial distance, which stops the velocity profile from dieing down at large $r$. However note that, the linear behavior of radion mass is only the leading order behaviour. If we had kept higher order terms, we would have corrections over and above the linear term, leading to, 
\begin{align}\label{Actual}
M_{\Phi}(r)=\frac{8\pi G\ell}{\kappa ^{2}}M_{B}\frac{r}{r_{0}}+C_{1}r^{\ell _{1}}+C_{2}r^{\ell _{2}}
\end{align}
where $C_{1}$ and $C_{2}$ are constants depending on $\kappa ^{2}/\ell$ and $\ell _{1}, \ell _{2}$ both are strictly less than unity. Given the mass profile, the corresponding velocity profile can be obtained by dividing the mass profile by $r$ and some suitable numerical factor. The coefficients and powers of the velocity profile (and hence the mass profile) can be determined by fitting the velocity profile with the observed one.
\begin{figure*}
\begin{center}
\includegraphics[scale=0.3]{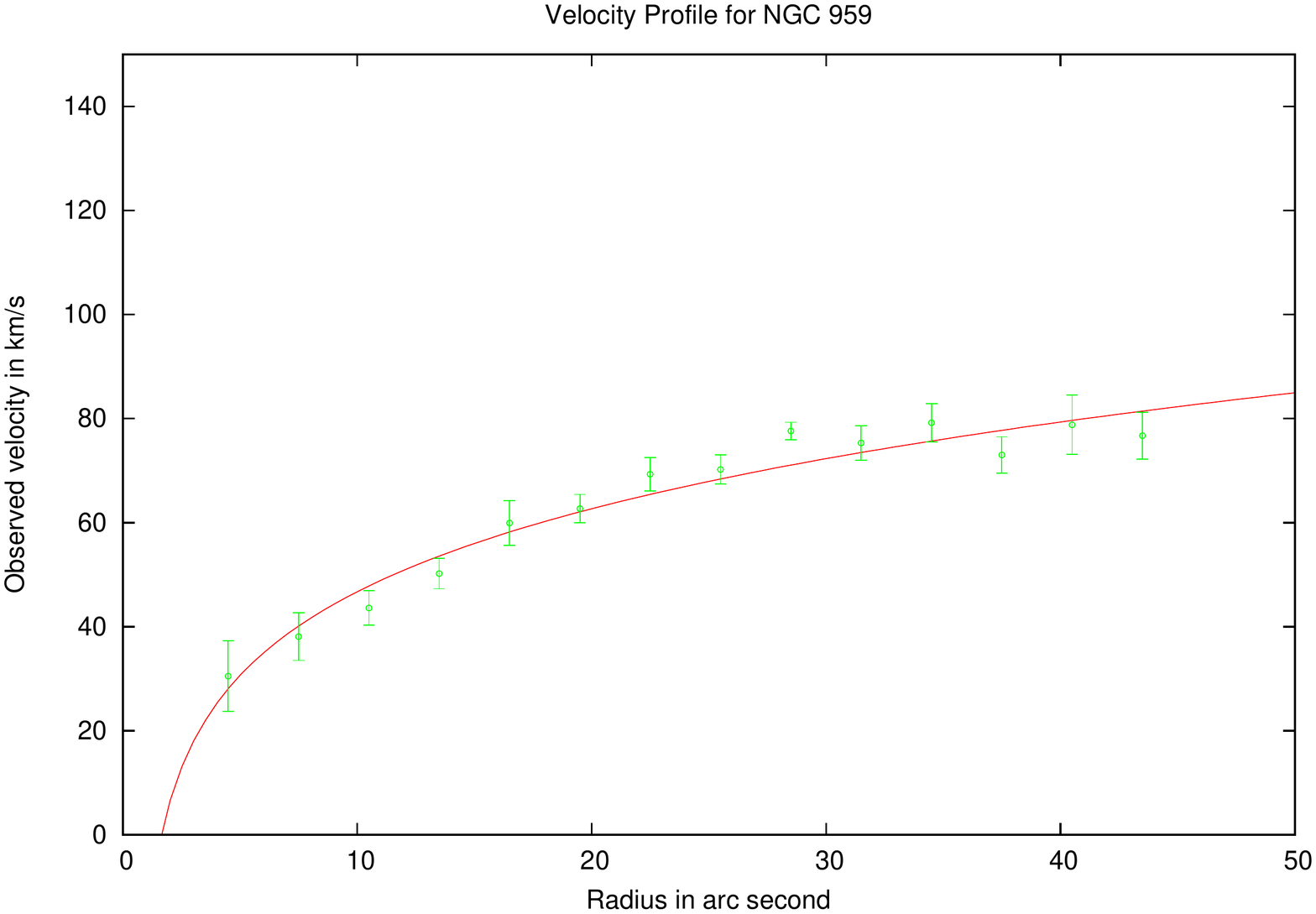}~~
\includegraphics[scale=0.3]{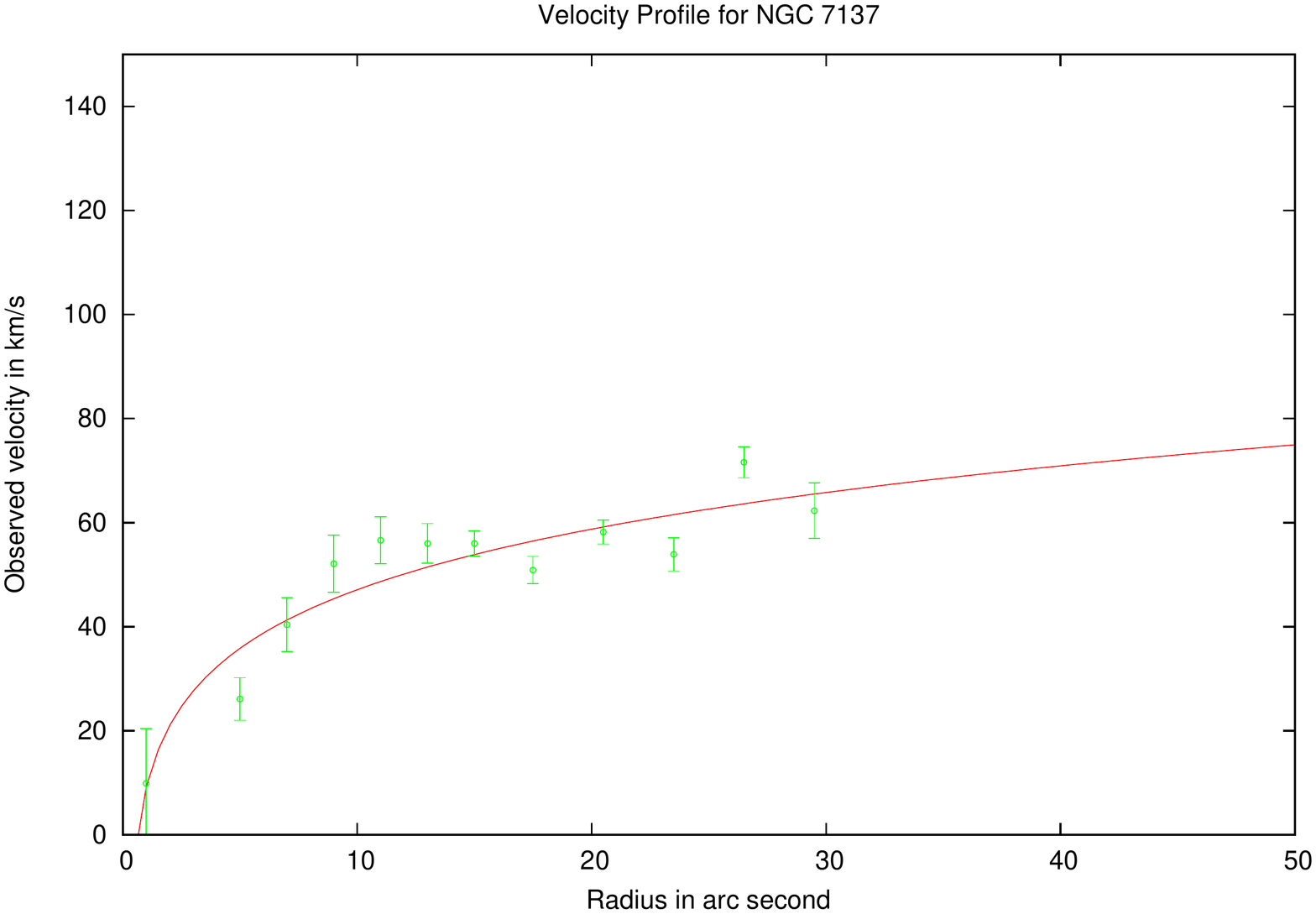}\\
\includegraphics[scale=0.3]{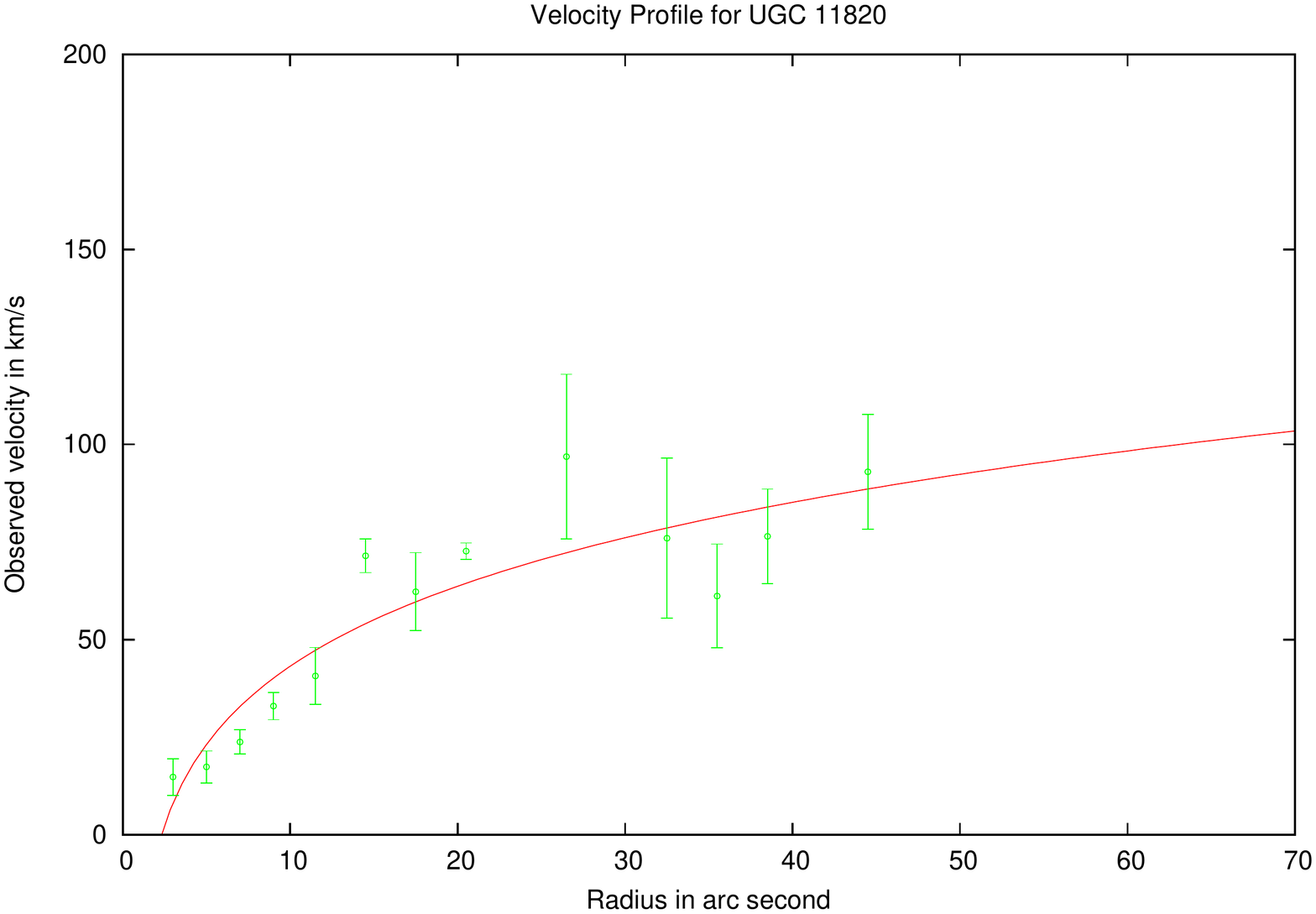}~~
\includegraphics[scale=0.3]{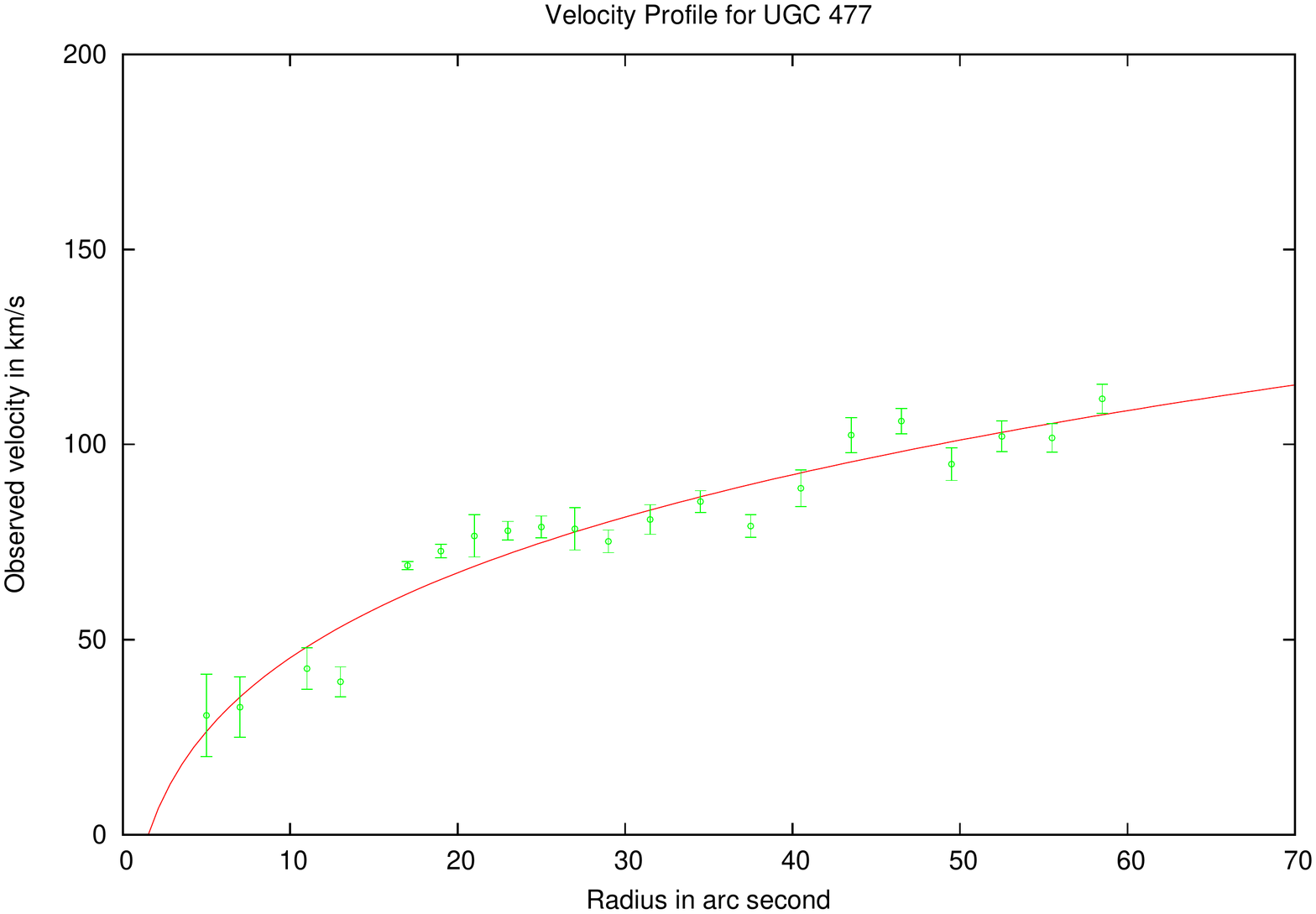}\\

\caption{Best fit curves for four chosen low surface brightness galaxies, NGC 959, NGC 7137, UGC 11820, UGC 477, respectively \cite{KuziodeNaray:2007qi,KuziodeNaray:2006wh}. On the vertical axis we have plotted observed velocity in km/s and the horizontal axis illustrates the radius measured in arc second. Good fit shows that the assumption of spherical symmetry is a good one, also the fact that baryonic matter plus radion field explains the galactic rotation curves fairly well. It also depicts the need for the sub-leading terms in \ref{Actual},}
\label{fig_01}
\end{center}
\end{figure*}
We should emphasize that the linear term alone cannot lead to good fit, its effect is to make the velocity profile flat at large distances. Thus at smaller distances the additional correction terms in \ref{Actual} are absolutely essential. Hence the effect of radion field can only be felt at large distances, preventing the velocity profile from decaying and the sub-leading factors in \ref{Actual} are important for matching with experimental data. In particular, from \ref{fig_01} it turns out that all the four curves are consistent with the following choices of the power law behavior: $\ell _{1}\simeq 0.1$ and $\ell _{2}\simeq 0.4$ respectively. The coefficients $C_{1}$ and $C_{2}$ turn out to have the following numerical estimates: $C_{1}=-25.56\pm4.3$ and $C_{2}=1.75\pm 0.08$ respectively. Thus at small enough values of $r$ the dominant contribution comes from the term $C_{1}r^{\ell _{1}}$, while for somewhat larger values of $r$, $C_{2}r^{\ell _{2}}$ dominates. Finally at large 
values of $r$ the linear term, i.e., the contribution from radion field becomes dominating, leading to flat velocity profile for the galaxies. Hence the correction terms are of quite significance in order to obtain a good fit with observational data.
\section{Application to other scenarios}\label{App_Scen}

In this work we have used a two brane model with brane separation being represented by the radion field $\Phi$. We have also assumed that our universe corresponds to the visible brane. In such a setup the effective gravitational field equations on the brane, written in a spherically symmetric context, depends on the radion field and its derivatives. Use of collisionless Boltzmann equation leads to the result:  \emph{virial mass of the galaxy clusters scales linearly with radial distance}. Thus without any dark matter we can reproduce the virial mass of galaxy clusters by invoking extra dimensions.

However in order to become a realistic model we should apply our results to other situations and look for consistency. There are mainly three issues which we want to address --- (i) advantage over other modified gravity models, (ii) reproducing the correct cosmology and (iii) connection with local gravity tests, in particular fifth force proposal. We address all these issues below.
\begin{itemize}

\item   In present day particle physics an important and long standing problem is the gauge hierarchy problem, which originates due to the large energy separation between the weak scale and Planck scale. In our model the branes are separated by a distance $d$, such that the energy scale on our universe gets suppressed by $M_{\rm vis}\sim M_{\rm Pl}e^{-2kd}$, with $k$ being related to brane tension. Thus a proper choice of $k$ (such that $kd\sim 10$ ) leads to $M_{\rm vis}\sim M_{\rm weak}$ and hence solves the hierarchy problem. Along with the missing mass problem, i.e., producing linear Virial mass our model has the potential of resolving the gauge hierarchy problem as well. This is a major advantage over modified gravity models, where the modifications in gravitational field equations are due to modifying the action for gravity. These models, though can explain the missing mass problem, usually does not address the gauge hierarchy problem.

\item  The next hurdle comes from local gravity tests. This should place some constraints on the behaviour of the radion field. The analysis using a spherically symmetric metric ansatz has been performed in \cite{deRisi:2012af} assuming dark matter to be a perfect fluid which is a perturbation over the Schwarzschild solution. We can repeat the same analysis with our radion field mass function, which is a perturbation over the vacuum Schwarzschild solution. We then can compute the correction to the perihelion precession of mercury due to dark matter which leads to the following constraint on the bulk curvature radius \cite{deRisi:2012af,Frere:2007pi,Bovy:2012tw}
\begin{equation}
\frac{2\ell (3-\beta)(\beta -2)}{\kappa ^{2}M_{\odot}}a(1-e^{2})\leq \frac{10^{-5}}{36^{2}\pi}\frac{T_{M}}{T_{E}}\Delta \delta \phi
\end{equation}
where $\Delta \delta \phi =0.004\pm 0.0006$ arc second per century, corresponds to excess in the perihelion precession of Mercury \cite{Fienga:2011qh}. Among others $a$ is the semi-major axis, $e$ stands for eccentricity, $T_{M}$ and $T_{E}$ are periods of revolution of Mercury and Earth respectively. 

\item Let us now briefly comment on the relation between existence of fifth force and dark matter. In all these models the generic feature corresponds to the existence of a scalar field which couples to dark matter and in turn couples weakly (or strongly) to standard model particles \cite{Gradwohl:1992ue,Bean:2008ac,Carroll:2008ub}. In our model this feature comes quite naturally, since the radion field $\Phi$ which plays the role of dark matter can also be thought of as a scalar field, couples to standard model particles through the matter energy momentum tensor with coupling parameter $\sim \kappa ^{2}/\ell(3+2\omega)^{-1}$. Thus effectively we require a fifth force to accommodate modifications of gravity at small scales. There exist stringent constraints on the fifth force from various experimental and observational results (see for example \cite{Adelberger:2003zx,Long:2002wn,Bertotti:2003rm}). We can apply these constraints on the fifth force for scalar tensor theories of gravity and that 
leads to the following bound on the composite object: $(\kappa ^{2}/12\pi G \ell)(1+\Phi)<2.5\times 10^{-5}$. Hence for compatibility of the radion field presented in this work with fifth force constraints, the bulk curvature $\ell$, the bulk gravitational constant $\kappa ^{2}$ and radion field must satisfy the above mentioned inequality.

\item Finally we address some cosmological implications of our work. In cosmology one averages over all the matter contributions at the scale of galaxy clusters and assumes all the matter components to be perfect fluid. The same applies to our model as well, in which the effect of radion field $\Phi$ at the galactic scale is to generate an effective dark matter density profile, with a given mass function. Since the mass function obeys the observed dark matter profile, therefore on the average in the cosmological scale it reproduces the standard dark matter content and hence the standard cosmological models. 

\end{itemize}
Thus the radion field model proposed in this work not only matches with the virial mass profile of galaxy clusters but also fits well into other scenarios. The model has the advantage over other alternative gravity models since it can also address the hierarchy problem by exponentially suppressing Planck scale on our universe. Secondly, local gravity tests and fifth force phenomenology provides constraints on bulk curvature radius which is consistent with virial mass profile. Still, the results can change depending on the stabilization mechanism for radion field, which would be an interesting future avenue to explore.
\section{Discussion}

Brane world models can address some of the long standing puzzles in theoretical physics, namely --- (a) the hierarchy problem and (b) the cosmological constant problem. To solve the hierarchy problem we need two branes, with warped five dimensional geometry such that energy scale on the visible brane gets suppressed exponentially leading to TeV scale physics. For the cosmological constant brane tension plays a crucial role. Two brane models naturally inherit an additional field, the separation between the branes (known as the radion field). Radion field is also very important in both macroscopic and microscopic physics, for it can have possible signatures in inflationary scenario \cite{Csaki:1999mp,Csaki:1999jh,Chakraborty:2013ipa}, black hole physics \cite{Kar:2015lma,Kar:2015fva}, collider searches \cite{Anand:2014vqa}, etc. Along with the gauge hierarchy and cosmological constant problem, another very important problem in physics, is the missing mass problem. This appears since baryonic and 
virial mass of a galaxy cluster do not coincide. In this work using a two brane setup we have shown that, along with gauge hierarchy and cosmological constant problem, this model is also capable of addressing the missing mass problem through the kinematics of the brane separation, i.e., radion field. Due to the presence of this additional field, the gravitational field equations on the brane gets modified and yields additional correction terms on top of Einstein's field equations. By considering relativistic Boltzmann equation we have derived the virial mass of galaxy clusters, which depends on an effective additional mass constructed out of radion field. Moreover these correction terms modifies the structure of gravity and hence the motion under its influence at large distance, thereby producing a linear increase in the virial mass of the galaxy clusters. This in turn leads to the appropriate velocity law for galaxies within a galaxy cluster, solving the missing mass problem. 
\section*{Acknowledgements}

S.C. thanks IACS, India for warm hospitality; a part of this work was completed there during a visit. He also thanks CSIR, Government of India, for providing a SPM fellowship. 
\appendix
\labelformat{section}{Appendix #1} 
\labelformat{subsection}{Appendix #1}
\section{Appendix: Derivations of Various Expressions Used in Text}\label{APP_01}

In this appendix, we summarize derivations of important expressions presented in the main text. We hope this will be helpful to clarify the important algebraic steps to arrive at various results in the main body of this paper.

We start with various derivatives of the scalar field $\Phi$ in this spherically symmetric coordinate:
\begin{align}
D^{2}\Phi &=\frac{1}{\sqrt{-g}}\partial _{\mu}\left(\sqrt{-g}g^{\mu \nu}\partial _{\nu}\Phi \right)
\nonumber
\\
&=\frac{1}{\exp [(\nu+\lambda)/2]r^{2}\sin \theta}\partial _{r}\left[\exp[(\nu -\lambda)/2]r^{2}\partial _{r}\Phi \right]
\nonumber
\\
&=e^{-\lambda}\partial _{r}^{2}\Phi +\frac{2}{r}e^{-\lambda}\partial _{r}\Phi +e^{-\lambda} \left(\frac{\nu '-\lambda '}{2}\right)\partial _{r}\Phi
\\
\left(D\Phi \right)^{2}&=g^{rr}\partial _{r}\Phi \partial _{r}\Phi =e^{-\lambda}\left(\partial _{r}\Phi \right)^{2}
\\
D_{r}D^{r}\Phi &=g^{rr}\left(\partial _{r}^{2}\Phi -\Gamma ^{r}_{rr}\partial _{r}\Phi \right)
=e^{-\lambda}\left(\partial _{r}^{2}\Phi -\frac{\lambda '}{2}\partial _{r}\Phi \right)
\\
D^{t}D_{t}\Phi &=-e^{-\lambda}\frac{\nu '}{2}\partial _{r}\Phi
\\
D^{\theta}D_{\theta}\Phi &=D^{\phi}D_{\phi}\Phi =\frac{1}{r}e^{-\lambda}\partial _{r}\Phi
\end{align}
Using which the field equation for $\Phi$ takes the following form:
\begin{align}
D_{\mu}D^{\mu}\Phi =\frac{\kappa ^{2}}{\ell}\frac{1+\Phi}{3}T^{(\rm{vis})}+\frac{1}{2(1+\Phi)}D_{\mu}\Phi D^{\mu}\Phi
\end{align}
Let us now turn our attention to the gravitational field equations. We will start with the temporal component such that:
\begin{align}
G^{t}_{t}&=\frac{\kappa ^{2}}{\ell}\frac{1}{\Phi}T^{t}_{t}+\frac{1}{\Phi}\left(D_{t}D^{t}\Phi -D^{2}\Phi \right)
+\frac{3}{4}\frac{1}{\Phi (1+\Phi)} \left(D\Phi \right)^{2}
\nonumber
\\
&=-\frac{\kappa ^{2}}{\ell}\frac{\rho +\Lambda}{\Phi}-e^{-\lambda}\frac{\nu '}{2}\frac{\partial _{r}\Phi}{\Phi}-\frac{\kappa ^{2}}{\ell}\frac{1+\Phi}{3\Phi}T-\frac{1}{2\Phi(1+\Phi)}\left(D\Phi \right)^{2}+\frac{3}{4}\frac{1}{\Phi (1+\Phi)} \left(D\Phi \right)^{2}
\nonumber
\\
&=-\frac{\kappa ^{2}}{\ell}\frac{\rho +\Lambda}{\Phi}-e^{-\lambda}\frac{\nu '}{2}\frac{\Phi '}{\Phi}-\frac{\kappa ^{2}}{\ell}\frac{1+\Phi}{3\Phi}T+\frac{e^{-\lambda}\Phi '^{2}}{4\Phi(1+\Phi)}
\end{align}
Then the radial component:
\begin{align}
G^{r}_{r}&=\frac{\kappa ^{2}}{\ell}\frac{1}{\Phi}T^{r}_{r}+\frac{1}{\Phi}\left(D_{r}D^{r}\Phi -D^{2}\Phi \right)-\frac{3}{2}\frac{1}{\Phi (1+\Phi)}D^{r}\Phi D_{r}\Phi+\frac{3}{4}\frac{1}{\Phi (1+\Phi)} \left(D\Phi \right)^{2}
\nonumber
\\
&=\frac{\kappa ^{2}}{\ell}\frac{p-\Lambda}{\Phi}+e^{-\lambda}\left(\frac{\Phi ''}{\Phi} -\frac{\lambda '}{2}\frac{\Phi '}{\Phi} \right)-e^{-\lambda}\frac{\Phi ''}{\Phi} -\frac{2}{r}e^{-\lambda}\frac{\Phi '}{\Phi} -e^{-\lambda} \left(\frac{\nu '-\lambda '}{2}\right)\frac{\Phi '}{\Phi}
\nonumber
\\
&-\frac{3}{2}\frac{\Phi '^{2}}{\Phi (1+\Phi)}e^{-\lambda}+\frac{3}{4}\frac{\Phi '^{2}}{\Phi (1+\Phi)} e^{-\lambda}
\nonumber
\\
&=\frac{\kappa ^{2}}{\ell}\frac{p-\Lambda}{\Phi}-\frac{2}{r}e^{-\lambda}\frac{\Phi '}{\Phi}-e^{-\lambda} \frac{\nu '}{2}\frac{\Phi '}{\Phi}-\frac{3}{4}\frac{\Phi '^{2}}{\Phi (1+\Phi)}e^{-\lambda}
\end{align}
and finally the transverse part yields:
\begin{align}
G^{\theta}_{\theta}&=G^{\phi}_{\phi}=\frac{\kappa ^{2}}{\ell \Phi}T^{\theta}_{\theta}+\frac{1}{\Phi}\left(D_{\theta}D^{\theta}\Phi -D^{2}\Phi \right)
+\frac{3}{4}\frac{1}{\Phi (1+\Phi)} \left(D\Phi \right)^{2}
\nonumber
\\
&=\frac{\kappa ^{2}}{\ell \Phi}\left(p_{\perp}-\Lambda \right)+\frac{1}{r}e^{-\lambda}\frac{\Phi '}{\Phi}-\frac{\kappa ^{2}}{3\ell}\frac{1+\Phi}{\Phi}T-\frac{1}{2}\frac{1}{\Phi (1+\Phi)} \left(D\Phi \right)^{2}+\frac{3}{4}\frac{1}{\Phi (1+\Phi)} \left(D\Phi \right)^{2}
\nonumber
\\
&=\frac{\kappa ^{2}}{\ell \Phi}\left(p_{\perp}-\Lambda \right)+\frac{1}{r}e^{-\lambda}\frac{\Phi '}{\Phi}-\frac{\kappa ^{2}}{3\ell}\frac{1+\Phi}{\Phi}T+\frac{1}{4}\frac{e^{-\lambda}\Phi '^{2}}{\Phi (1+\Phi)} 
\end{align}
Addition of these three equations and assuming the system to be non-relativistic, \ref{GB_Eq02} leads to, 
\begin{align}
\frac{\nu ''}{2}+\frac{\nu '}{r}&=\frac{\kappa ^{2}}{2\ell \Phi}\rho-\frac{1}{4}\frac{\Phi '^{2}}{\Phi (1+\Phi)}-\frac{\kappa ^{2}}{6\ell}\frac{1+\Phi}{\Phi}\left(-\rho -4\Lambda \right)-\frac{\kappa ^{2}}{\ell \Phi}\Lambda
\nonumber
\\
&=\frac{\kappa ^{2}}{6\ell}\frac{4+\Phi}{\Phi}\rho +\frac{\kappa ^{2}}{3\ell}\frac{2\Phi -1}{\Phi}\Lambda-\frac{1}{4}\frac{\Phi '^{2}}{\Phi (1+\Phi)}
\nonumber
\\
&=\frac{\kappa ^{2}}{6\ell}\rho +\frac{2\kappa ^{2}}{3\ell}\Lambda-\frac{1}{4}\frac{\Phi '^{2}}{\Phi (1+\Phi)}+\frac{2\kappa ^{2}\rho}{3\Phi \ell}-\frac{\kappa ^{2}\Lambda}{3\ell \Phi}
\end{align}
Multiplying \ref{GB_Eq05} with $4\pi r\rho (r)$ and integrating we obtain:
\begin{align}\label{GB_App}
\frac{1}{2}\int 4\pi \rho(r)r^{3}\nu 'dr&=\frac{\kappa ^{2}}{24\pi \ell}\int 4\pi \rho (r)rM(r)dr
+\frac{2\kappa ^{2}\Lambda}{3\ell}\int 4\pi \rho (r)r\frac{r^{3}}{3}dr +\frac{\kappa ^{2}}{4\pi \ell}\int 4\pi \rho (r)rM_{\Phi}dr
\nonumber
\\
&=\frac{\kappa ^{2}}{24\pi \ell}\int \frac{M(r)}{r}dM(r)
+\frac{2\kappa ^{2}\Lambda}{9\ell}\int r^{2}dM(r) +\frac{\kappa ^{2}}{4\pi \ell}\int \frac{M_{\Phi}}{r}dM(r)
\end{align}
These are the expressions used in main text. We also need to define the following objects:
\begin{align}
\Omega =-\int \frac{GM}{r}dM;\qquad \Omega _{\Phi}=-\int \frac{GM_{\Phi}}{r}dM;\qquad I=\int r^{2}dM
\end{align}
Then \ref{GB_App} takes the following form:
\begin{align}
2K+\frac{\kappa ^{2}}{24\pi G\ell}\Omega +\frac{\kappa ^{2}}{4\pi G\ell}\Omega _{\Phi} -\frac{2\kappa ^{2}\Lambda}{9\ell}I=0
\end{align}
Let us introduce three radius $R_{V}$, $R_{I}$ and $R_{\Phi}$ as:
\begin{align}
R_{V}&=\frac{M^{2}}{\int \frac{M}{r}dM};\qquad R_{I}^{2}=\frac{\int r^{2}dM}{M};\qquad R_{\Phi}=\frac{M_{\Phi}^{2}}{\int \frac{M_{\Phi}}{r}dM}
\end{align}
Then the above defined objects, namely, $\Omega$, $I$ and $\Omega _{\Phi}$ reduces to,
\begin{align}
\Omega =-\frac{GM^{2}}{R_{V}};\qquad I=MR_{I}^{2};\qquad \Omega _{\Phi}=-\frac{GM_{\Phi}^{2}}{R_{\Phi}}
\end{align}
On using these relations in the expression for Kinetic energy K, finally leads to
\begin{align}
\frac{GM_{V}^{2}}{R_{V}}+\frac{\kappa ^{2}}{24\pi G\ell}\left(-\frac{GM^{2}}{R_{V}}\right)-\frac{2\kappa ^{2}\Lambda}{9\ell}\left(MR_{I}^{2}\right)+\frac{\kappa ^{2}}{4\pi G\ell}\left(-\frac{GM_{\Phi}^{2}}{R_{\Phi}}\right)=0
\end{align}
The above expression on rearrangement leads to \ref{GB_Final}. Now we need to solve for $\Phi$, which satisfies the following differential equation,
\begin{align}
\Phi ''+\frac{2}{r}\Phi ' +\left(\frac{\nu '-\lambda '}{2}\right)\Phi '=\frac{\kappa ^{2}}{3\ell}\left(1+\Phi \right)\left(-\rho-4\rho_{0} \right)+\frac{\Phi '^{2}}{2(1+\Phi)}
\end{align}
which under integration leads to,
\begin{align}
r^{2}\frac{\Phi '}{1+\Phi}=-\frac{\kappa ^{2}}{3\ell}\int r^{2}\rho (r)dr=-\frac{\kappa ^{2}}{12\pi \ell}M(r)
\end{align}
Integrating once again we will immediately obtain \ref{Rad_Final}.

\bibliography{Brane,Gravity_1_full,Gravity_2_partial,Galaxy_Rotation,Cosmology,My_References}

\bibliographystyle{./utphys1}
\end{document}